\documentclass[aps,showpacs,floatfix,twocolumn]{revtex4}
\pdfoutput=1
\usepackage{graphics, color}
\usepackage{graphicx}
\usepackage{amssymb}
\usepackage{amsmath}
\usepackage{subfigure} 
\usepackage[hang,small,justification=raggedright]{caption}
\usepackage{epsfig}
\def\gsim{\, \rlap{$>$}{\lower 1.1ex\hbox{$\sim$}}\,}
\def\lsim{\, \rlap{$<$}{\lower 1.1ex\hbox{$\sim$}}\,}
\def\p{\partial}

\newcommand{\be}{\begin{equation}}
\newcommand{\ee}{\end{equation}}
\newcommand{\bal}{\begin{align}}
\newcommand{\eal}{\end{align}}
\newcommand{\ben}{\begin{equation*}}
\newcommand{\een}{\end{equation*}}
\newcommand{\bea}{\begin{eqnarray}}
\newcommand{\eea}{\end{eqnarray}}
\newcommand{\bean}{\begin{eqnarray*}}
\newcommand{\eean}{\end{eqnarray*}}
\newcommand{\bes}{\begin{subequations}}
\newcommand{\ees}{\end{subequations}}

\begin{document}
\title{Holographic Description of Finite Size Effects in Strongly Coupled Superconductors}
\author{Antonio M. Garc\'{\i}a-Garc\'{\i}a}
\affiliation{Cavendish Laboratory, University of Cambridge, Thomson Avenue, Cambridge, CB3 0HE, UK}
\author{Jorge E. Santos, and Benson Way}
\affiliation{Department of Physics, University of California, Santa Barbara, CA 93106-4030}
\begin{abstract}
Despite its fundamental and practical interest, the understanding of mesoscopic effects in strongly coupled superconductors is still limited. Here we address this problem by studying holographic superconductivity in a disk and a strip of typical size $\ell$.  For $\ell < \ell_c$, where $\ell_c$ depends on the chemical potential and temperature, we have found that the order parameter vanishes. The superconductor-metal transition at $\ell = \ell_c$ is controlled by mean-field critical exponents which suggests that quantum and thermal fluctuations induced by finite size effects are suppressed in holographic superconductors. Intriguingly, the effective interactions that bind the order parameter increases as $\ell$ decreases. Most of these results are consistent with experimental observations in Pb nanograins at low temperature and qualitatively different from the ones expected in a weakly coupled superconductor.
\end{abstract}
\maketitle
The AdS/CFT correspondence \cite{adscft}, also known as gauge/gravity duality or holography, states that under certain conditions, strongly coupled gauge theories in $d$ dimensions are dual to a classical gravity theory in Anti de Sitter (AdS) space in $d+1$ dimensions. Information about the strongly coupled field theory can thus be obtained by simply solving the equations of classical Einstein gravity with appropriate matter content. In recent times, the potential applications of gauge/gravity duality  beyond high energy physics has become a major focus of research in theoretical physics. Progress in the context of superconductivity \cite{hhh} has been especially encouraging (see e.g. \cite{review} for a review).  For example, there are holographic descriptions of different types of strongly coupled superconductors \cite{hhh,us},  superconductor-insulator transitions \cite{takayanagi}, non-Fermi liquids \cite{nfl}, tunnel junctions \cite{hojo} and field theories with more general boundaries \cite{taka1}. 

One of the downsides of the holographic approach is that the dual field theory that describes the condensed matter system is not explicitly known. Therefore a key problem in the field is to identify exactly the range of applicability of the holographic approach in realistic condensed matter systems. This paper is a small step in this direction.  We study finite size (FS) effects in holographic superconductors (in a strip and a disc), and compare these results with condensed matter predictions \cite{teo1,teo2,teo3,expcm,nmat}. The main conclusion is that there are striking similarities between our results and those corresponding to strongly coupled superconductors, such as Pb, at low temperature.
 
\begin{figure*}
\includegraphics[width=0.65\columnwidth,clip,angle=0]{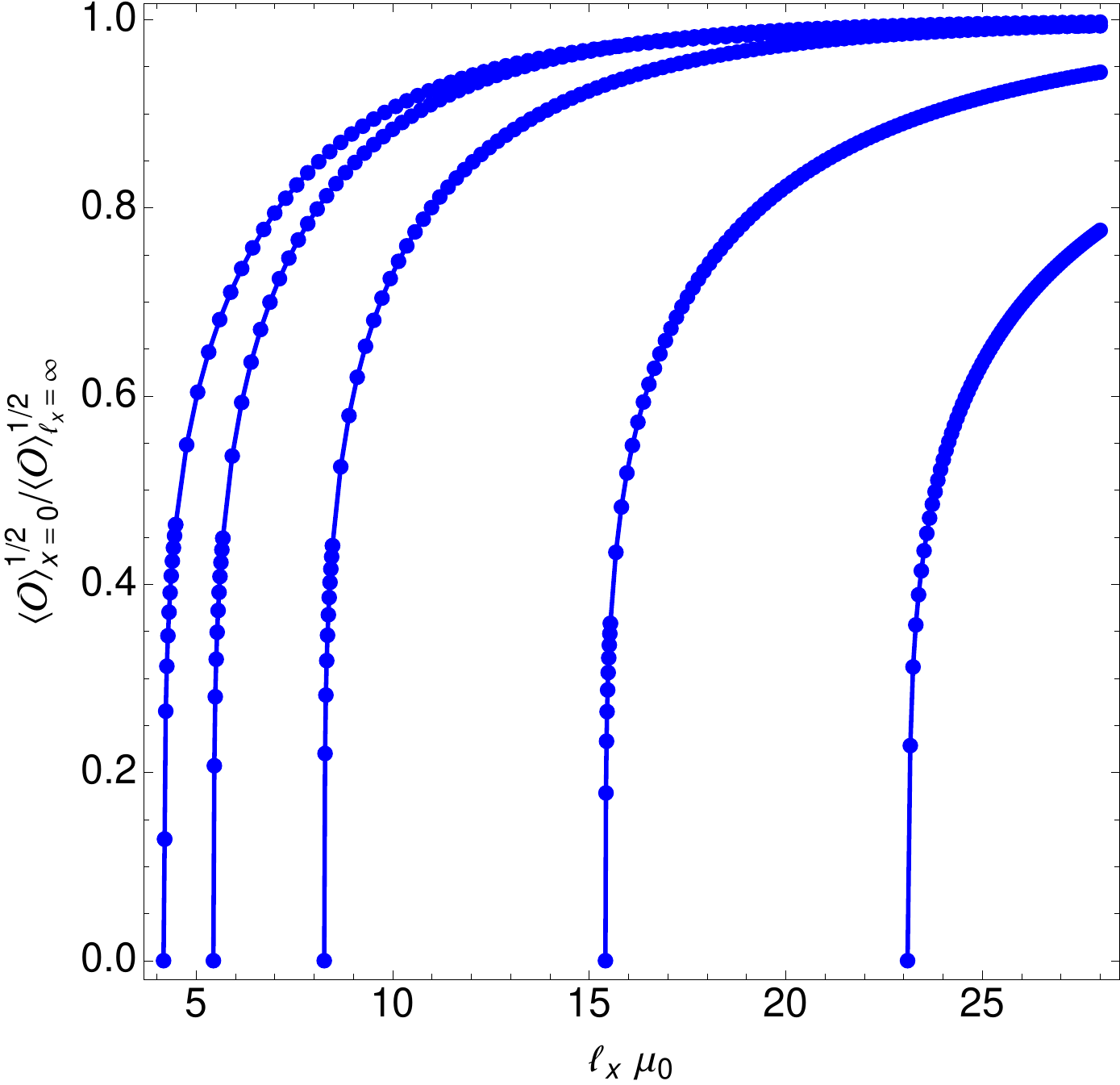}
\qquad\qquad
\includegraphics[width=0.65\columnwidth,clip,angle=0]{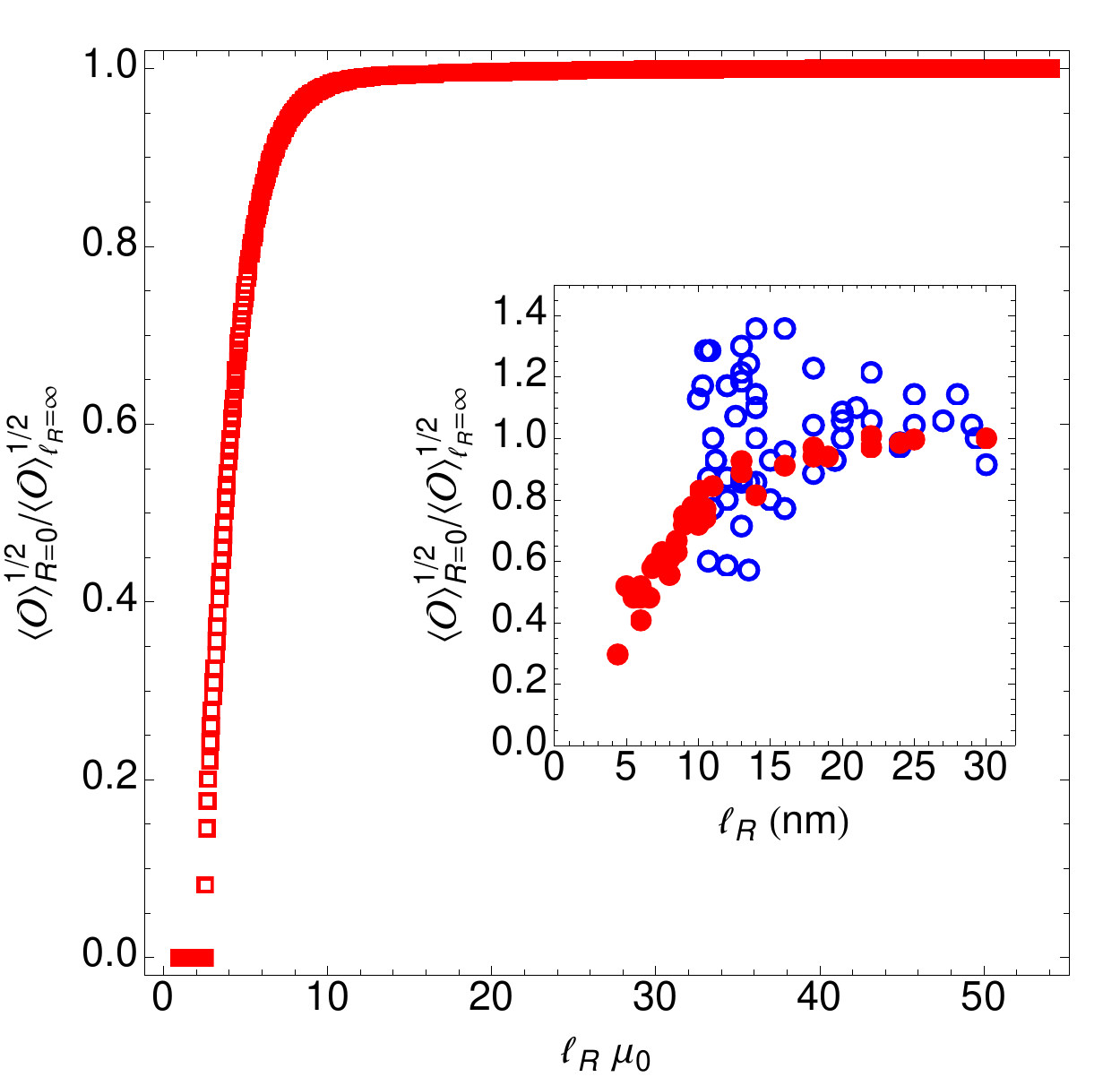}
\caption{Left: Normalized order parameter versus strip width $\ell_x$ in units of $\mu_0$ for (left to right)  $T/T_c =$ $0.3$, $0.52$, $0.73$, $0.9$ and $0.95$. Right: the same for a disk of radius $\ell_R$ and $T= 0.5T_c$. Inset: Experimental value of the gap, obtained by STM techniques, in single isolated Pb (red solid circles) and Sn (blue circles) hemispherical nanograins \cite{nmat} for $T \approx 0.2T_c$ as a function of the grain radius $\ell_R$.}
\label{fig1}
\end{figure*}

For the sake of completeness we give a brief overview of previous research on FS effects in superconductivity. On the experimental side, recent advances in the growth and control of nanostructures \cite{expcm} have made it possible to study the evolution of superconductivity in single isolated nanograins of different materials as the system size $\ell$ is gradually reduced.  In weakly coupled superconductors \cite{nmat} such as Al or Sn (see Fig. \ref{fig1}), the order parameter has an intricate oscillating pattern as a function of $\ell$ in the region $\ell \ll {\rm max}\{\xi,l \}$ where $\xi$ and $l$ are the superconducting and single particle coherence length, respectively. The amplitude of the oscillations increases substantially at low temperature and in systems with substantial level degeneracies.  This is a typical coherence effect similar to shell effects in nuclear physics.  

By contrast, oscillations are suppressed \cite{nmat} in strongly coupled superconductors such as Pb and, to a good approximation, the gap decreases monotonically with $\ell$. For $\ell < \ell_c$ the order parameter vanishes. There are no experimental results about FS effects in cuprates.  However, it is expected that similar results hold since shorter $\xi$ and/or the breaking of the Fermi liquid theory is expected to suppress coherence effects. On the theoretical side, it is expected that, at least in weakly coupled superconductors, thermal \cite{teo2} and quantum fluctuations \cite{teo3} induced by FS effects smooth out the transition when the order parameter becomes comparable to the mean level spacing. It was shown in \cite{ambe} that if quantum fluctuations are suppressed due to strong inelastic scattering or a very short $\xi$, the order parameter vanishes for $\ell \leq \ell_c$. 

This manuscript is organized as follows.  We introduce the holographic model and highlight the main steps leading to the calculation of the order parameter for different sizes and geometries.  Then we discuss the results of this calculation and its relation to condensed matter systems.  
 
\emph{The Model} -- 
The gravity theory is defined by the action,
\be\label{action}
S=\int d^4x\;\sqrt{-g}\left[R+\frac{6}{L^2}-\frac{1}{4}F_{\mu\nu}F^{\mu\nu}-|D\psi|^2-m^2|\psi|^2\right],
\ee
where $F=dA$ and $D=\nabla-iqA$. This simple model has been shown to lead to superconductivity \cite{hhh} in the dual field theory.  Superconductivity in this model is related to the spontaneous breaking of a U(1) symmetry for sufficiently low temperatures. We will restrict our analysis to the probe approximation in which the backreaction of these fields on the metric is ignored. That is,  we rescale $\psi=\tilde\psi/q,$ $A=\tilde A/q$ and take $q\rightarrow\infty$, keeping $\tilde\psi$ and $\tilde A$ fixed. Therefore, we can fix the metric background to be the AdS planar Schwarzschild black hole:
\begin{align}
ds^2&=-f(r)dt^2+\frac{dr^2}{f(r)}+r^2(dx^2+dy^2)\label{cart}\\ 
&=-f(r)dt^2+\frac{dr^2}{f(r)}+r^2(dR^2+R^2d\theta^2)\label{polar}\\ 
&\qquad f(r)=\frac{r^2}{L^2}\left(1-\frac{r_0^3}{r^3}\right) \;,\nonumber
\end{align}
where $r_0$ is the black hole horizon radius, and $L$ is the AdS length scale.  We will use \eqref{cart} to model the strip and \eqref{polar} to model the disc.  The black hole temperature is given by,
$
T=\frac{3r_0}{4\pi L^2}\;.
$
In order to proceed, we consider solutions of the form
\be
\tilde\psi=|\psi|\;,\qquad \tilde A=A_t\;dt\;,
\ee
where $|\psi|$, and $A_t$ are positive functions of $x$ and $r$ for the strip and $R$ and $r$ for the disc.  The equations of motion using \eqref{cart} are,
\begin{widetext}
\bea
\p_{r}^2|\psi|+\frac{1}{r^2f}\p_{x}^2|\psi|+\left(\frac{f'}{f} + \frac{2}{r}\right)\p_r|\psi|+\frac{1}{f}\left(\frac{A_t^2}{f}-m^2\right)|\psi|=0\;,\label{psi}\\
\p_{r}^2A_t+\frac{1}{r^2f}\p_{x}^2A_t+\frac{2}{r}\p_r A_t-\frac{2|\psi|^2}{f}A_t=0\;.\label{At}
\label{tot_equations}
\eea
\end{widetext}
We now discuss boundary conditions.  For convenience, we set $L=1$ and $m^2=-2$.  At the horizon $r=r_0$, regularity requires $A_t=0$.  This, together with the equations of motion, fixes the horizon boundary condition on the remaining function $|\psi|$.  At $x=\pm \infty$ or $R=\infty$, we require that the functions approach the normal-phase homogeneous solutions with $|\psi|=0$.  

Near $r=\infty$, the fields have the following form:
\bea
|\psi|=\frac{\psi^{(1)}}{r}+\frac{\psi^{(2)}}{r^2}+O\left(\frac{1}{r^3}\right)\\
A_t =\mu-\frac{\rho}{r}+O\left(\frac{1}{r^2}\right)\;.
\eea
Each of the coefficients in $r$ are functions of either $x$ or $R$.  The $\psi^{(1)}$ and $\psi^{(2)}$ terms are both normalizable, so we have a choice of boundary conditions.  Here, we only consider the case $\psi^{(1)}=0.$  Via gauge/gravity duality,  $\psi^{(2)}$ then gives the expectation value of a dimension two operator in the boundary field theory. Therefore the order parameter of the holographic superconductor is simply
\bea
\langle \mathcal{O}\rangle= \sqrt{2}\psi^{(2)}.
\label{op}
\eea
The main aim of this manuscript is to compute $\langle \mathcal{O}\rangle$.  

The quantities $\mu$ and $\rho$ are interpreted in the boundary field theory as the chemical potential and the charge density, respectively.  In order to model finite size effects, we introduce a spatially dependent $\mu(\vec x)$ \cite{Keranen:2009ss,Flauger:2010tv}. A similar method was recently used in \cite{hojo} to model a Josephson junction.  In this case, we choose a profile for $\mu(\vec x)$ so that $\mu(\vec x)$ remains approximately constant in some region near the origin, and then drops sharply around an $\ell$ corresponding to the boundary of the disk or strip. In a certain range of temperatures, this will yield a superconducting region with the shape of a disk or a strip and a normal metal region everywhere else. For numerical reasons, the drop of $\mu(\vec x)$ around the disk or strip boundary, though sharp, has to be smooth. A profile for the strip that meets these requirements is
$$
\mu(x)=\mu_0\left[\frac{1-\epsilon+\epsilon\cosh\left(\frac{2x}{\sigma}\right)+\cosh\left(\frac{\ell_x}{\sigma}\right)}{\cosh\left(\frac{2x}{\sigma}\right)+\cosh\left(\frac{\ell_x}{\sigma}\right)}\right]\;,
$$
where $\mu_0$ is the chemical potential at the origin. 
To obtain a profile for the disc, we relabel $x\rightarrow R$ and $\ell_x\rightarrow2\ell_R$ where $\ell_x$ is the width of the strip and $\ell_R$ the radius of the disc.  The critical temperature of the superconductor $T_c(\ell\mu_0)$ with  $\ell = \ell_R, \ell_x$ is proportional to the chemical potential. For $\ell \to \infty$,  $T_c \equiv T_c(\infty) \approx0.0588\mu_0$.  

The quantities $\sigma$ and $\epsilon$ control the steepness and depth of our profile, respectively.  We choose a steepness $\sigma \gg \ell$ so that $\mu(x)$ is quite flat inside the system.  The effective critical temperature far from the origin is given by $T_{\infty}=\epsilon\,T_c$.   We choose $\epsilon$ so that the region outside the superconductor remains in the normal phase for the range of temperatures considered.  For this work, we employ $\sigma/\ell=0.1$ and $\epsilon=0.3$.

We now attempt to solve the above system of coupled nonlinear partial differential equations.  We follow the same numerical procedure outlined in \cite{hojo}.  As in \cite{hojo}, we use scaling symmetries to  fix the chemical potential $\mu_0$.  All lengths are then expressed in terms of the dimensionless quantity $\ell \mu_0$.  

\emph{Results} -- 
In this section we study the dependence of the order parameter (\ref{op}) and the critical temperature on the size and shape (disk or strip) of the system.  Our main motivation is to gain a deeper understanding of holographic superconductivity by comparing it with those found in real materials. First we study the evolution of the order parameter as a function of the typical system size $\ell \equiv \ell_x,\ell_R$ for different temperatures. In order to prevent the region outside the disk or the strip from becoming superconducting, and to avoid a breakdown of the probe approximation, the minimum temperature that we investigate is $0.3T_c$. The main findings of this analysis, depicted in Fig. \ref{fig1}, are as follows: (a) the order parameter vanishes for $\ell < \ell_c(T,\mu)$ so the instability towards superconductivity only occurs for sufficiently large systems; this is typical of mean-field treatments of superconductivity \cite{ambe}; (b) the transition is controlled by mean-field critical exponents $\langle\mathcal{O}(\ell\mu_0)\rangle^{1/2} \propto \sqrt{\ell - \ell_c}$; we note that the observation of a true phase transition, even in a FS system, is a direct consequence of the large $N$ approximation involved in the holographic calculation \cite{witten}; (c) in the range of sizes investigated, the order parameter decreases monotonically as the system size decreases; this is in clear contrast with weakly coupled superconductors such as Sn \cite{nmat} where large oscillations of the order parameter, related to a longer $\xi$ \cite{teo1}, have been observed experimentally (see inset Fig. \ref{fig1}); (d) as it is observed in the inset of Fig. \ref{fig1}, the predictions of FS effects in holographic superconductivity are strikingly similar to recent 
experimental results for Pb, a strongly coupled superconductor.  

Finally, we note that Pb is a type I superconductor while holographic superconductors are type II \cite{hhh}. FS effects in cuprates and other type II superconductors, though not yet known experimentally, are expected to be even closer to our results because in these materials $\xi$ is shorter which suppresses coherence effects. 
\begin{figure}
\includegraphics[width=0.65\columnwidth,clip,angle=0]{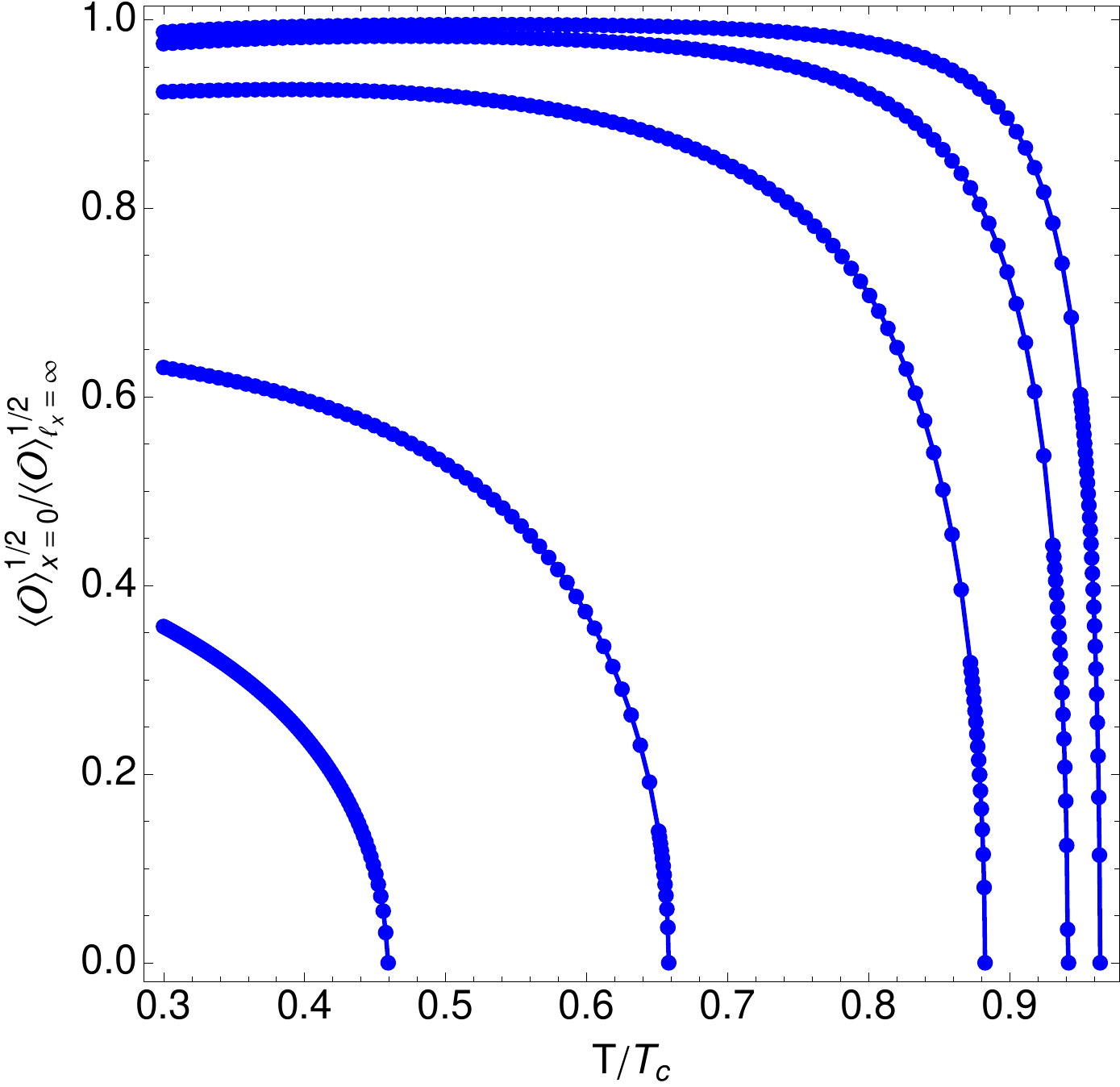}
\caption{Normalized condensate from Eq. (\ref{op}) as a function of temperature $T$ for different strip sizes, from right to left, $\ell_x\mu_0 = 28, 21, 14, 7,5$. Similar results (not shown) are obtained for the disk.}
\label{fig2}
\end{figure}  
 \begin{figure}
\includegraphics[width=0.65\columnwidth,clip,angle=0]{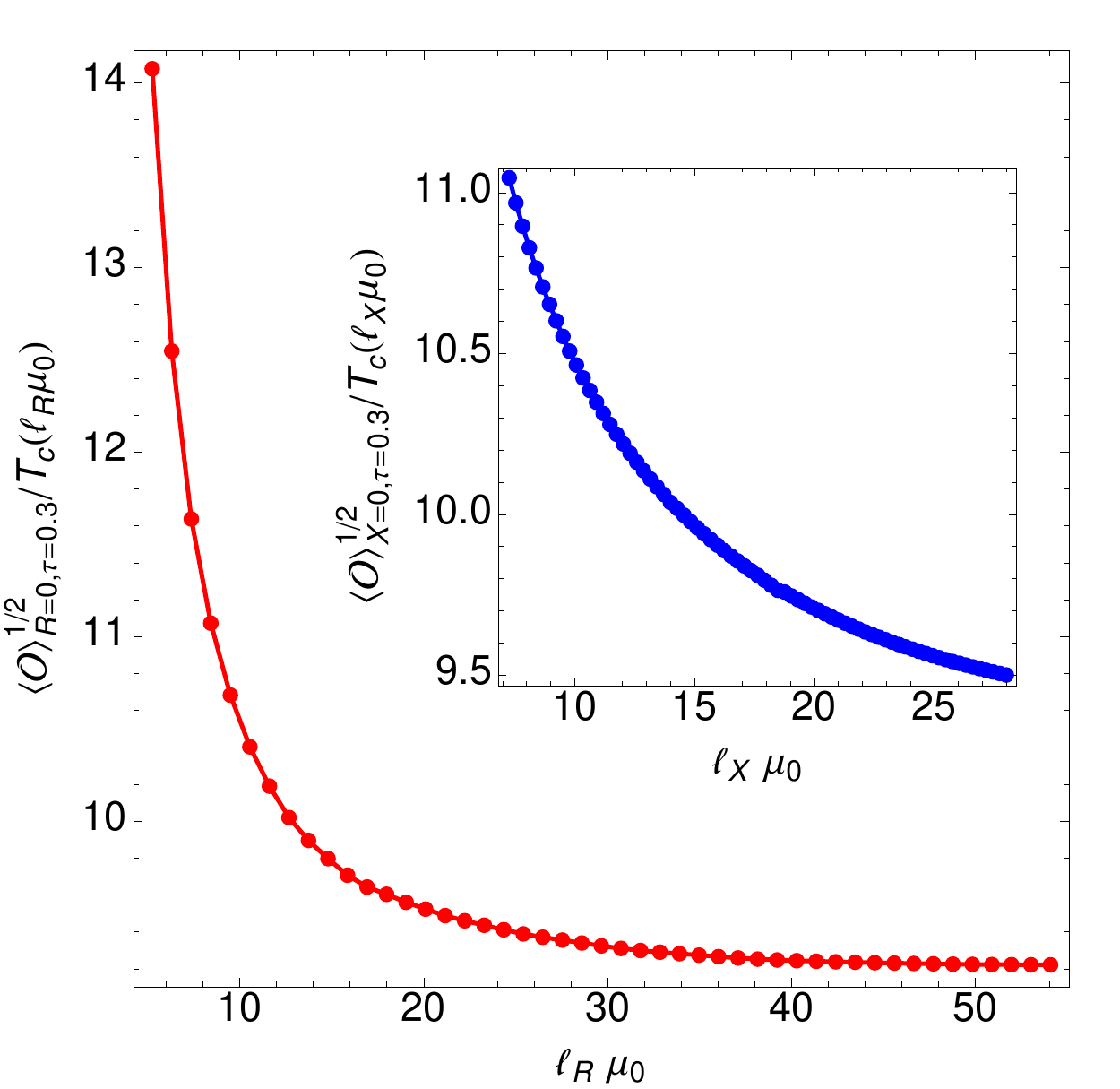}
\caption{$\langle\mathcal{O}(\ell_R\mu_0)\rangle^{1/2}/T_c(\ell_R\mu_0)$ as a function of the radius $\ell_R$. $\langle\mathcal{O}(\ell_R\mu_0)\rangle$ is  evaluated at $\tau \equiv T/T_c =0.3$. Inset: The same for a strip of width $\ell_x$. 
The ratio increases as the size decreases which suggests that FS effects enhance superconductivity.}
\label{fig3}
\end{figure}  

We now study FS effects on $T_c(\ell)$. Ginzburg-Landau theory predicts that there should be a region around $T_c(\ell)$ where deviations from mean-field predictions due to thermal fluctuations are expected. The main effect of thermal fluctuations is to smooth out the transition so that the amplitude of the order parameter is not zero even for $T > T_c(\ell)$. Assuming that $\xi < \ell$, the interval of temperatures in which thermal fluctuations are important is given by $\frac{\delta T}{T_c} \propto \sqrt{\frac{\delta}{T_c}}$, where $\delta T \equiv T - T_c$ and $\delta$ is the mean level spacing. Therefore, thermal fluctuations become more important as the system size is decreased. However, the temperature dependence of the order parameter of the holographic superconductor, shown in Fig. \ref{fig2}, follows the mean field prediction for all system sizes and shapes, even for $T \approx T_c$.  This is again a consequence of the large $N$ limit \cite{witten}.  The dependence on $\delta$ above can give us a clue to interpret the large $N$ limit in condensed matter systems. We speculate that in this context large $N$ is equivalent to adding a strong source of inelastic scattering or dissipation which effectively make $\delta$ vanish. Although in Pb nanograins deviations from the mean-field prediction have been observed for $\ell \sim 10$nm  \cite{nmat} we expect that our results will hold for cuprates and other high $T_c$ superconductors for which interactions are stronger than in Pb. 

Finally, we investigate the strength of the interactions that binds the condensate as a function of $\ell$. This will reveal interesting features of the physics of holographic superconductors.  A useful quantity is the dimensionless ratio $\sqrt{\langle \mathcal{O}(\ell) \rangle}/T_c(\ell)$ at low temperature.  In our calculation, we compute $T_c(\ell)$ by fitting curves like those in Fig. \ref{fig2} to mean field predictions, and compute $\sqrt{\langle \mathcal{O}(\ell) \rangle}$ at $\tau=T/T_c=0.3$, which is a good approximation to the zero temperature solution if $\ell$ is not too small.  

In a weakly-coupled superconductor such a Al or Sn, this quantity is a universal number $\approx 3.52$ that it is independent of the coupling constant. This is directly related to the fact that the effective coupling that binds the condensate is roughly the same at zero and at finite temperature. The vanishing of the order parameter at $T_c$ occurs because at finite temperature quasiparticles gradually occupy the states available for pairing to the point that at $T=T_c$ long range order is no longer possible. 
 
In strongly coupled superconductors the ratio is typically larger and increases as the strength of the interaction at zero temperature becomes stronger. For conventional superconductors such as Pb the ratio is larger because the phonon-electron interaction decreases with temperature as a consequence of retardation effects related to the phonon dynamics. Even in conventional superconductors, there is no clear understanding of the size dependence of this ratio.  Surface phonons, which become more important as the system size decreases, tend to enhance the gap at zero temperature without modifying $T_c$ substantially \cite{ginzburg}.  
At the same time Coulomb interactions become stronger for smaller systems because screening become less effective. It is therefore natural to expect that in general the ratio $\sqrt{\langle O(\ell) \rangle}/T_c(\ell)$ is size dependent, though there is not yet any conclusive experimental evidence. 

For a holographic superconductor, see Fig. \ref{fig3}, we observe a monotonic increase of the ratio as $\ell$ decreases. This suggests that holographic superconductivity is enhanced by FS effects. It is tempting to speculate that this enhancement is caused by the gradual loss of screening of the force in the boundary theory that leads to superconductivity. This is reminiscent of the cuprates where a reduction of doping in the overdoped phase, increases Coulomb interactions, and lead to a higher $T_c$.
 
In conclusion, we have investigated FS effects in holographic superconductors with the aim to reveal the potential of this approach for modelling realistic systems. We have found that, in agreement with recent results in Pb nanograins, the order parameter decreases monotonically with the system size until it vanishes in a mean-field fashion for sufficiently small sizes. Thermal
fluctuations are also suppressed as a consequence of the large $N$ approximation which suppress FS coherence effects. Finally we have shown holographic superconductivity is enhanced as $\ell$ decreases. This has been related to the gradual reduction of screening of the force that binds the condensate. 

AMG acknowledges support from EPSRC under Grant No. EP/I004637/1, FCT under Project PTDC FIS/111348/2009 and from a Marie Curie International Reintegration Grant PIRG07-GA-2010-268172. JS and BW acknowledge support from NSF Grant No. PHY08-55415, and thank G. Horowitz for helpful discussions.

\end{document}